\documentstyle[12pt,epsf]{article}
\input epsf
\def\beq{\begin{eqnarray}}   \def\eeq{\end{eqnarray}}

\begin{document}
\begin{flushright}
NYU-TH/00/04/01 \\
April 25, 2000
\end{flushright}

\vspace{0.1in}
\begin{center}
\bigskip\bigskip
{\large \bf 4D Gravity on a Brane in 5D  Minkowski Space} 

\vspace{0.5in}      

{Gia Dvali, Gregory Gabadadze, Massimo Porrati}
\vspace{0.1in}

{\baselineskip=14pt \it 
Department of Physics, New York University,  
New York, NY 10003 } 
\vspace{0.2in}
\end{center}

\vspace{0.9cm}
\begin{center}
{\bf Abstract}
\end{center} 
\vspace{0.1in}

We suggest a mechanism by which four-dimensional Newtonian gravity
emerges on a 3-brane in 5D Minkowski space with an infinite size 
extra dimension.  The worldvolume  theory gives rise to  
the correct 4D potential at short distances whereas  at 
large distances the potential is that of a 5D theory.
We discuss some phenomenological issues in this framework.

\newpage

\vspace{0.2in}
{\bf 1. Introduction}
\vspace{0.1in} \\

The observed weakness of gravity may be due to the fact that we live
on a brane embedded in space with large extra dimensions \cite {ADD}.
The correct 4D gravity can be reproduced at large distances 
due to the {\it finite~ volume} of extra space. 
This is usually achieved by compactifying extra space.  
Alternatively, this can be  obtained by  
keeping extra space  {\it uncompactified} 
but {\it warped} as in the 
scenario of  \cite {RandallSundrum}, where
the size  $L$ of extra space  
is still finite ($L=2\int_0^{\infty} \sqrt{g}dy<\infty$)~. 

In this work  we shall discuss the fate of 
4D gravity in theories with {\it infinite}  size  {\it flat}  
extra dimensions. 
These models may shed new light on
supersymmetry breaking and the cosmological constant problem
\cite {DGP1,Witten}, since they
make compatible unbroken bulk supersymmetry with fermi-bose
non-degeneracy on a brane \cite {DvaliShifman,DGP1,Witten} 
(for related discussions, see \cite {Dvali}).
A first  example with an infinite size extra dimension
and  the correct 4D Newtonian potential  
was proposed in \cite {GRS1} as a generalization of
the scenario of \cite {RandallSundrum}. 
However, the model of \cite {GRS1} has ghosts
\footnote{
The 4D Newtonian force  in \cite {GRS1} is mediated by a  
resonance graviton \cite {GRS1,Csaki1,DGP1} which has extra 
polarization degrees of freedom \cite {DGP1}.   
The brane construction of \cite {GRS1}
violates the null-energy  condition \cite {Witten},
and, as a result, it has a state with negative norm, a {\it ghost}
\cite {DGP2,Kogan,Kang,Riccardo}. 
A brane bending term \cite {Csaki2} which emerges 
in the traceless-transverse gauge, is a manifestation of a ghost 
\cite{DGP2}. In the harmonic gauge the ghost can be identified with the 
radion field which has a negative kinetic term \cite {Riccardo}.
This ghost can cancel the unwanted polarizations of 
a resonance graviton \cite {Csaki2,GRS2,Riccardo}. However, the presence of
a propagating ghost signals the 
inconsistency \cite {DGP2,Riccardo} of the model of Ref. \cite{GRS1}.}.  

In the present paper we suggest a general mechanism by which  
4D Newtonian gravity may  be generated  
on any static 3-brane embedded in 5D Minkowski space.  
We show that the 4D scalar curvature term  in 
the worldvolume brane action
\beq
M^2_{P}~\int d^4x ~\sqrt {|g|}~{R}~,
\nonumber
\eeq
can be responsible for the correct 4D Newtonian interaction on a brane, 
despite of the fact that gravity propagates in  5D Minkowski space.  
Such a term is compatible with all the symmetries 
left unbroken on a  brane and can be included in the theory.
Moreover, if it is absent in a classical theory,
it may  be generated on a brane by quantum corrections.
It turns out that inclusion of such a term automatically generates
the $1/r$ gravitational potential  at short distances for 
the sources localized on a  brane. As a result, an  observer on a  brane
will see  correct Newtonian gravity despite of the fact that
gravity propagates in  extra space which is flat and has the infinite extent.
We should note  here that  the modification of
gravity on a brane due to this effect  
in weakly coupled string theory 
would take place at very short, 
phenomenologically unacceptable distances.
Nevertheless, it is interesting to study this mechanism 
in a low-energy field theory framework.

The paper is organized as follows. In section 2 we present  a   
5D model with an infinite size extra dimension
and discuss  mechanisms  by which 4D Newtonian gravity 
on a brane can be obtained from higher dimensional theory.
In section 3 we study a prototype model which involves only a scalar
field  and calculate the potential which is produced by mediation of this
scalar. At  short distances the potential has 4D nature, while 
it has 5D behavior at large distances.  
In section 4 we study the 4D graviton 
propagator in the infinite extra dimensional model. 
As in the case with a scalar, gravitons mediate the 4D
Newton potential at small distances with a crossover to 
the 5D potential at large distances. 
Discussions of some phenomenological issues are given in section 5.

\vspace{0.2in}
{\bf 2. A  Model of 4D Gravity in 5D Minkowski Space}
\vspace{0.1in} \\

We start with a $D=(4+1)$ dimensional theory. Let us suppose 
there is  a 3-brane 
embedded in $5$-dimensional space-time. We assume that
this is a zero-tension brane in $5$-dimensions. 
The four coordinates of our world are  $x_\mu,~\mu=0,1,2,3$;
the extra coordinate will be denoted by  $y$. 
Capital letters and subscripts will be used for 
5D quantities  ($A,B,C=0,1,2,3,5$);  
the metric convention is mostly negative.

Let us consider the action:
\beq
S~=~ M^3~\int d^5X ~\sqrt {G}~{\cal R}_{(5)}~+~
M^2_{P}~\int d^4x ~\sqrt {|g|}~{R}~,
\label{1}
\eeq
where $M$ stands for the 5D Planck mass, and $M_P$ 
is the 4D Planck mass; as they stand in (\ref {1}) $M$ and $M_P$  
are independent (in general they can be dependent).  
$G_{AB}(X) \equiv G_{AB}(x,y)$ denotes
a 5D metric for which the 5D Ricci scalar is
${\cal R}_{(5)}$. The brane is located at $y=0$.
The induced metric on the  brane is denoted by
\beq
g_{\mu\nu}(x)~\equiv~G_{\mu\nu} (x,~y=0)~. 
\label{4Dg}
\eeq
The 4D Ricci scalar for $g_{\mu\nu}(x)$ is $R=R(x)$. 
Possible additional terms 
of the corresponding SUGRA and/or
matter fields confined to a  brane  are omitted in 
Eq. (\ref {1}) for simplicity. 

In the limit $M\rightarrow 0$ with finite $M_P$ 
the action (\ref {1}) 
describes 4D gravity  on a brane. On the other hand, in the 
limit $M_P\rightarrow 0$  with finite $M$ it describes 5D bulk gravity.
In what follows we study 4D gravitational interactions 
on a brane when both $M$ and $M_P$ are finite.  

Before we turn to these discussions 
let us try to understand a possible  origin
of the action (\ref {1}). 

The first possibility is related to the fact that 
in certain cases (to be specified below)  
the localized matter on a brane can induce via loop corrections a 
4D kinetic term for gravitons. To demonstrate this,  
suppose there are matter fields confined to  a brane.
Thus, the matter energy-momentum tensor can be  written as follows: 
\beq
T_{AB}~=~\left (
\begin{tabular} {c c}
$T_{\mu\nu}(x)~ \delta (y)$ & ~~0 \\
0              & ~~0    \\
\end{tabular}
\right )~.
\eeq
As a result, the interaction Lagrangian of localized 
matter with  5D metric fluctuations $h_{AB}(x,y)\equiv G_{AB}(x,y)-\eta_{AB}$, 
reduces to the following expression: 
\beq
{\cal L}_{\rm int} ~=~ \int dy~h^{\mu\nu}(x,y) ~T_{\mu\nu}(x)~\delta(y)~=~
h^{\mu\nu}(x,0) ~T_{\mu\nu}(x)~,
\label{int4D}
\eeq
where the 4D induced metric $g_{\mu\nu}(x)=\eta_{\mu\nu}+h_{\mu\nu}$ 
is defined as in (\ref {4Dg}).
Due to this interaction, a 
4D kinetic term can be generated  for 
$g_{\mu\nu}(x)$  in the full quantum theory. 
For instance, the diagram  of Fig. 1 with massive 
scalars \cite {Capper},  or fermions \cite {Adler,Zee}
running in the loop  would  induce the following  4D term
in the low-energy action:
\beq
\int d^4x~dy~\delta(y)~\sqrt{|g|}~R~.
\label{ind}
\eeq 
\begin{figure}
\centerline{\epsfbox{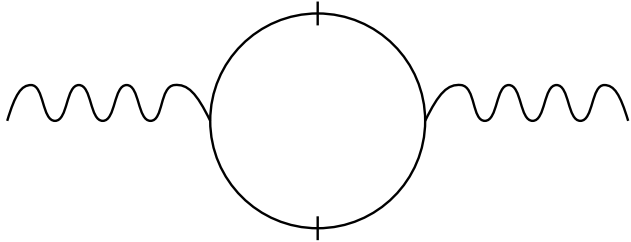}}
\epsfysize=6cm
\caption{\small The  one-loop diagram generating the 
Ricci scalar  for 4D graviton. Wave lines denote gravitons, 
solid lines denote massive scalars/fermions. Vertical short lines on 
scalar/fermion propagators indicate that they are massive.}
\label{fig1}
\end{figure} 
The corresponding induced gravitational constant will be determined by
a correlation function of the world-volume matter theory\footnote{For 
instance, in the 4D framework of Refs. 
\cite {Adler,Zee}:
$$
{M_P^2} ~=~ {i\over 96} \int d^4x ~x^2~ \left 
\{ \langle TS(x) S(0)\rangle -\langle S\rangle ^2  \right \}~,
\label{G}
$$
where 
$S(x) \equiv T^{\mu}_{\mu}$ is the trace of the energy-momentum tensor
of 4D states running in the loop.}. The magnitude of this constant 
depends on a worldvolume theory at hand and is vanishing  
in conformaly invariant models, or nonzero 
if conformal invariance is broken 
(for detailed discussions see \cite {Adler,Zee,Khuri}).
We will not attempt to  discuss these model dependent features 
here, we  rather assume that
the worldvolume theory is such that the second term in (\ref {1}) 
is generated with a proper sing and magnitude.
We also neglect  the 
induced 4D cosmological constant (which renormalizes the brane tension), 
$\Lambda =\langle 0|T^{\mu}_{\mu}|0\rangle$, as well as 
higher derivative terms ${\cal O}(R^2)$ which can be   
generated  in this case as well. 
The induced 4D cosmological constant, in general, should be 
canceled by the brane tension and/or by the vacuum energy due to light 
matter fields on a brane. Without SUSY this 
is a usual fine tuning.  However, if the 
bulk is supersymmetric (even though SUSY is broken on a brane) 
it might require the exact cancellation of the total  4D cosmological 
constant in a model which at large distances (low energies) 
becomes effectively 5-dimensional. 
This certainly needs further separate investigation. Here we
assume that  the total 4D cosmological constant is  zero.   
 
The second way to think of the origin of the 4D term in (\ref {1})
is to imagine that 5D gravity is coupled to a certain 5D scalar field
$\phi $. Suppose that in addition to usual terms there 
are terms in which  $\phi$ couples to  
the five-dimensional Ricci scalar.
Furthermore, suppose that the scalar field potential is 
such that there is a kink type solution to its  
equations of motion:
$\phi_{\rm classical}=v{\rm tanh}(vy)~$.
This background, due to the presence of higher derivative interactions
might produce the 4D  Ricci scalar which would be  peaked around 
the point $y=0$. In the limit  of a zero width for the kink 
this term can be approximated by  (\ref {ind}). 
A prototype example with scalars will be presented  in the next  section.
Then we turn to the consideration of gravitons. 

\vspace{0.2in} 
{\bf 3. A Simple Example with Scalars}
\vspace{0.1in} \\

In this section we study  an instructive example 
which involves a single 5D scalar field. 
We write the action of the model as follows:
\beq
S=M^3\int d^4xdy ~\partial_A\Phi(x,y) \partial^A\Phi(x,y) ~+~
M_P^2\int d^4xdy ~\delta(y) \partial_\mu \Phi(x,0)\partial^\mu\Phi(x,0)~.
\label{scalars}
\eeq 
For the purpose of 
comparison with gravity we choose to work with the  
unconventionally normalized dimensionless scalar field $\Phi$.
Let us notice that this kind of action could 
arise if the 5D field $\Phi$ is placed in the background 
$\chi~=~v~{\rm tanh}(vy)~$, and  the interactions 
between these fields are taken as: $(\partial_A\Phi)^2+
(\partial_A\chi)^2(\partial_B\Phi)^2-(\partial_A\Phi \partial^A\chi)^2+...$.
This leads to the equation of motion for $\Phi$: $2\partial_A^2 \Phi +
(\chi ^{\prime})^2\partial_\mu^2 \Phi=0$. Here the prime denotes 
derivative with respect  to $y$. 
$(\chi ^{\prime})$ is a peaked function at $y=0$ point. 
Therefore, this equation  can be modeled by the action (\ref{scalars}).

Our goal is to  determine the distance dependence of interactions
which are mediated by this scalar
in a  4D worldvolume theory. For this 
we should find the corresponding retarded Green  function  
and calculate the potential. 
The classical equation for the Green  function  looks as follows:
\beq
\left ( M^3 \partial_A\partial^A~+~M_P^2~\delta(y)~ \partial_\mu\partial^\mu 
\right ) ~G_R(x,
y; 0,0)~=~\delta^{(4)}(x) \delta(y)~,
\label{green}
\eeq
where $G_R(x,y; 0,0)=0$ for $x_0<0$.
 
The potential mediated by the scalar $\Phi$ 
on the 4D worldvolume of the brane is determined as:
\beq
V(r)~=~\int G_R\left (t,{\overrightarrow x},y=0; 0,0,0\right 
) dt~,
\label{pot}
\eeq
where $r\equiv\sqrt{x_1^2+x_2^2+x_3^2}$. 
To find a solution of (\ref {green}) let us turn to 
Fourier-transformed quantities with respect to 
the worldvolume four-coordinates $x_\mu$:
\beq
G_R(x,y; 0,0)~\equiv~\int ~ {d^4p\over (2\pi)^4}~e^{ipx} ~{\tilde G}_R(p,y)~. 
\label{Fourie}
\eeq 
Turning to Euclidean space the equation (\ref {green})
takes the form:
\beq
\left (~ M^3(p^2-\partial_y^2)~+~M_P^2~ p^2 ~\delta(y) ~\right )~  
{\tilde G}_R(p,y)~=
~\delta(y)~. 
\label{mom}
\eeq
Here $p^2$ denotes the square of an Euclidean four-momentum.
The solution with appropriate boundary conditions 
takes the form:
\beq
{\tilde G}_R(p,y)~=~{1\over M_P^2p^2~+~2M^3p}~ {\rm exp} (-p|y|)~,
\label{sol1}
\eeq
where $p\equiv\sqrt {p^2}=\sqrt{p_4^2+p_1^2+p_2^2+p_3^2}$. 
Using this expression and Eq. (\ref {pot}) one finds the following
(properly normalized)
formula for the  potential mediated by a scalar in 4D brane worldvolume:
\beq
V(r)~=~-{1\over 8\pi^2 M_P^2}~{1 \over r}~\left \{ {\rm sin} 
\left ( {r\over r_0} \right ) ~{\rm Ci} \left ( {r\over r_0} \right )
~+~{1\over 2}  {\rm cos} 
\left ( {r\over r_0} \right ) \left 
[\pi~-~2 ~ {\rm Si} \left ( {r\over r_0} \right ) \right ]   \right \}~,
\label{V} 
\eeq
where $ {\rm Ci}(z) \equiv \gamma +{\rm ln}(z) +\int_0^z ({\rm cos}(t) -1)dt/t$,
$ {\rm Si}(z)\equiv \int_0^z {\rm sin}(t)dt/t$,  
$\gamma\simeq 0.577$  is  the Euler-Masceroni 
constant, and the distance scale $r_0$ is defined as follows:
\beq
r_0~\equiv ~{M_P^2\over 2 M^3}~.
\label{r0}
\eeq
It is useful to study  the short distance and long distance
behavior of this expression.

At short distances when $r<<r_0$ we find:
\beq
V(r)~\simeq~-{1\over 8\pi^2 M_P^2}~{1 \over r}~\left \{
{\pi\over 2} +\left [-1+\gamma+{\rm ln}\left ( {r\over r_0} \right ) 
\right ]\left ( {r\over r_0} \right )~+~{\cal O}(r^2)  
\right \}~.
\label{short}
\eeq
As we expected, at short distances the potential 
has the correct 4D Newtonian $1/r$ scaling. This is subsequently modified
by the logarithmic {\it repulsion} term in (\ref {short}). 
 
Let us turn now to the large distance behavior. Using (\ref {V})
we obtain for $r>>r_0$:
\beq
V(r)~\simeq~-{1\over 8\pi^2 M_P^2}~{1 \over r}~\left \{
{r_0\over r}~+~{\cal O} \left ( {1\over r^2} \right ) 
\right \}~.
\label{long}
\eeq
Thus, the long distance potential
scales as $1/r^2$ in accordance with laws of 5D theory.
These properties are similar to those obtained in the model of Ref. 
\cite {GRS1}.  

In the next section we consider the system defined in (\ref {1}).
We will show that the short and long distance behavior of
the Newtonian potential for (\ref {1}) 
is determined by the scalar example studied  in this section. 
 
\vspace{0.2in} 
{\bf 4. Gravitational potential}
\vspace{0.1in} \\

Based on the scalar field example discussed in the previous
section  we expect  that the system (\ref {1})
will produce the $1/r$ gravitational potential at short distances
and $1/r^2$ potential at large scales. Therefore, if $r_0$ in 
(\ref {r0}) is big enough there will be no contradictions  
with Newtonian predictions.
However, there is a subtlety when it comes to relativistic 
effects. This is  related to the structure of 
the graviton propagator. 
The tensor structure of a 4D massless graviton propagator
looks as follows (we omit   
momentum-dependent parts):
\beq
{1\over 2} \eta^{\mu\alpha}  \eta^{\nu\beta}~+~
{1\over 2}  \eta^{\mu\beta}  \eta^{\nu\alpha}~-~   
{1\over 2}  \eta^{\mu\nu}  \eta^{\alpha\beta}~,         
\label{4D}
\eeq  
while for a 4D massive case, or equivalently for a  5D
massless case, it takes the following form:
\beq
{1\over 2} \eta^{\mu\alpha}  \eta^{\nu\beta}~+~
{1\over 2}  \eta^{\mu\beta}  \eta^{\nu\alpha}~-~   
{1\over 3}  \eta^{\mu\nu}  \eta^{\alpha\beta}~.         
\label{5D}
\eeq  
The difference in the last coefficient ( $1/2$ versus 
$1/3$) is vital for description of observations. 
It was shown in Refs. 
\cite {Veltman,Zakharov} that no matter how small the graviton mass
is in 4D, due to the difference in the tensor structures in 
(\ref {4D}) and (\ref {5D}), 
predictions for bending of light are substantially  different 
in the two cases. Moreover,  
the structure (\ref {5D}) gives contradictions with observations.
Any theory with massive gravitons, no matter how light they are,
will have to face this  problem since  
there is a discontinuity in the limit when the graviton mass
is taken to zero. This can easily be understood in terms of degrees of 
freedom: a massive graviton  
has 5 degrees of freedom 3 of which couple to a 
conserved energy-momentum tensor. Thus, having the propagator 
as in (\ref {5D}) is equivalent of having a tensor-scalar
gravity from 4D point of view. This extra scalar polarization 
degree of freedom yields  
additional  attractive force.
We will continue this discussion in the next section
after we find out what is the correct tensor 
structure of the graviton propagator in  the model (\ref {1}). 

To find the propagator 
let us introduce the metric fluctuations:
\beq
G_{AB} ~=~ \eta_{AB} ~+~h_{AB}~.
\eeq
We choose the {\it harmonic~ gauge} in the bulk: 
\beq
\partial^A h_{AB}~ =~{1\over 2}~ \partial_B h^C_C.
\label{gauge}
\eeq 
It can be checked that the choice
\beq
h_{\mu 5} ~=~0,
\label{mu5}
\eeq
is consistent with the equations of motions for (\ref {1}).
Thus, the surviving components of $h_{AB}$ are 
$h_{\mu\nu}$ and $h_{55}$.
In this gauge the (55) component of Einstein's  equations yields:
\beq
\partial_\mu\partial^\mu ~h^\nu_\nu ~=~\partial_\mu\partial^\mu~ h^5_5~,
\eeq
which combined with the gauge fixing conditions (\ref {gauge}, 
\ref {mu5})  implies:
\beq
\partial_A\partial^A ~ h^\nu_\nu ~=~\partial_B\partial^B ~h^5_5~.
\label{55}
\eeq
Subscripts and superscripts in all these equations 
are raised and lowered  by a flat space  metric tensor.
Finally, we come to the ($\mu\nu$) components
of the Einstein equation for (\ref {1}). After some rearrangements 
it  takes the form:
\beq
\left ( M^3 \partial_A\partial^A~+~M_P^2 ~\delta(y)~\partial_\mu\partial^\mu 
\right )~h_{\mu\nu}(x,y) ~=~ \left \{ T_{\mu\nu} -{1\over 3}\eta_{\mu\nu} 
T^\alpha_\alpha   \right \}~\delta(y)~\nonumber \\
+~ M_P^2~\delta(y) ~\partial_\mu
\partial_\nu ~h^5_5~.  
\label{basic}
\eeq
This has a structure of a massive 4D graviton or, equivalently 
that of a massless 5D graviton, 
indicating that the tensor structure of the 
propagator  looks as in (\ref {5D}). 
In this respect, it is instructive to rewrite
the expression  (\ref {basic}) in the following form:
\beq
\left ( M^3 \partial_A\partial^A~+~M_P^2~\delta(y)~
\partial_\mu\partial^\mu 
\right )~h_{\mu\nu}(x,y) ~=~ \left \{ T_{\mu\nu} -{1\over 2}\eta_{\mu\nu}  
T^\alpha_\alpha   \right \}~\delta(y)~ \nonumber \\
-{1\over 2} 
~M^3 ~\eta_{\mu\nu} ~\partial_A\partial^A~h^\alpha_\alpha
~+~ M_P^2~\delta(y)~ \partial_\mu \partial_\nu ~h^5_5~.  
\label{basic1}
\eeq
Here the tensor structure on the r.h.s. is that
of a 4D massless graviton (\ref {4D}). However, there is an additional
contribution due to the trace part $h^\mu_\mu$ which is nonzero.  
Therefore, one is left with 
the theory of gravity which from the 4D point of view is mediated
by a graviton plus a scalar\footnote{The same 
result could be obtained by using the traceless-transverse gauge and 
taking into account the ``brane bending'' effects \cite 
{GarrigaTanaka,Giddings}. We would like to emphasize here 
that the change of the coefficient $1/3$ to the 
coefficient $1/2$ in  \cite {GarrigaTanaka,Giddings}
is due to the possibility to perform gauge 
transformations of $h_{\mu\nu}$ 
which generate terms proportional to $\eta_{\mu\nu}$.
This term appears due to the  nontrivial warp factor. 
In the present case the warp-factor is absent,
thus the terms proportional to $\eta_{\mu\nu}$ are 
absent in gauge transformations and
the change  of the coefficient $1/3$ into $1/2$ does not take place.}.

Let us now present the exact form for the graviton propagator 
of (\ref {1}). Turning to the Fourier images in the Euclidean space
as in the previous section we find:
\beq
{\tilde h}_{\mu\nu}(p, y=0)~{\tilde T}^{\mu\nu}(p)~=~
{  {\tilde T}^{\mu\nu}{\tilde T}_{\mu\nu}~-~{1\over 3} ~{\tilde T}^{\mu}_\mu
{\tilde T}^{\nu}_\nu \over M_P^2 p^2~+~2M^3 p }~.
\label{prop}
\eeq  
Here the tilde sign  denotes  the Fourier-transformed
quantities. 
Thus, the tensor structure of the  graviton propagator in 4D worldvolume 
theory  looks as follows:
\beq
D^{\mu\nu\alpha\beta}~=~        
{1\over 2} \eta^{\mu\alpha}  \eta^{\nu\beta}~+~
{1\over 2}  \eta^{\mu\beta}  \eta^{\nu\alpha}~-~   
{1\over 3}  \eta^{\mu\nu}  \eta^{\alpha\beta} ~+~{\cal O} (p).
\label{propagator}
\eeq
At short distances the potential scales as 
$1/r$ with the logarithmic corrections defined in (\ref {short}). 
On the other hand, at   
large distances  the $1/r^2$ 
behavior is recovered in (\ref {long}).
The tensor structure of the propagator is that 
of 4D tensor-scalar gravity.  
 
\vspace{0.2in} 
{\bf 5. Discussions and Outlook}
\vspace{0.1in} \\

So far we worried about the theoretical consistency of the 
model with an infinite size  extra dimension (\ref {1}).
In this section we address the issue
whether this model could be phenomenologically 
viable.  
There are two outstanding questions in this regard.
The first one deals with the extra degree of freedom
which shows up in the propagator  (\ref {propagator}):
one should wonder if it is possible to  
cancel it.

The second question concerns the  modification of the 
Newton law by logarithmic corrections (\ref {short}): 
what is the distance at which
these modifications will  be harmless?

We will discuss these two issues in turn.
Let us start with the extra degree of freedom. 
As we said above, there is additional attraction
in the theory due to the extra scalar mode.   
We  should look for  
some new states  which could  compensate for this extra attraction.
This can certainly be done by a ghost field which gives rise to 
repulsive force.  However, it is hard to make sense of a
theory with a manifest ghost. The ghost  could  be produced
as a final state in various processes  and  this would ruin the consistency 
of the model at hand. Thus, we  should  exclude the possibility of 
using a ghost and try to utilize some  other 
means. 

A possibility to compensate for additional 
attraction would be to use an exchange 
of a  vector particle. 
One could think of a scenario where all matter fields are given 
additional $U(1)$ charges so that the corresponding
gauge fields  lead to repulsive interactions.  Then one should tune the 
parameters of the model so that this repulsion cancels 
(at least partially) the additional attraction in (\ref {propagator}).  
Whether this scenario suffices a single extra $U(1)$ 
or needs a few of them (so that the possible equivalence principle 
violation is not observable) should be determined by 
separate phenomenological studies. It is also possible that there exists 
a more elegant solution to this problem.  

Let us now turn to the second issue, what is the 
crossover scale where gravity in (\ref {1}) changes its 
behavior? As we discussed in section 3 this scale is determined by the 
ratio $r_0=M_P^2/2M^3$. Taking the value of the four-dimensional Planck scale 
$M_P\simeq 10^{19}~{\rm GeV} $ and assuming that
the five dimensional Planck scale is in the ${\rm TeV}$
region, we find that $r_0\sim 10^{15}~{\rm cm}$.
This is precisely the size of the solar system.  
However, the crossover  scale should be much bigger than  
the solar system size. 
For instance, requiring that the logarithmic  corrections 
in (\ref {short}) give the effects which are small compared
to the General Relativity corrections
to the precession of the Mercury perihelion, one obtains that
the crossover scale should be about $10^8$ times bigger
then the solar system size.  This means that $M$ should be 
about $10^2-10^3$ times smaller
then the ${\rm TeV}$ scale\footnote{As we mentioned before,
in non-supersymmetric D-brane constructions within weakly 
coupled string theory induced 4D terms 
will modify gravity at distances less than the string scale.}.
It is not obvious at present whether such a low scale   
can be obtained from higher dimensional theories. 

In conclusion, we suggested  a class of models in which 
4D Newtonian gravity can emerge on a brane in 5D flat  space.  
A crucial  feature of these models is that 
the extra dimension is neither compact nor warped and 
its size is truly infinite.  
The modification of the Newton law takes place at large distances
in this framework.  In the minimal setup there are  
phenomenological subtleties. We outlined some possible ways
to circumvent them by adding extra fields. 
More detailed studies are
needed in order determine whether 
models with infinite size extra dimensions 
can be given a complete, phenomenologically viable  form.

We would like to emphasize that the similar  mechanism can be used for
``localization'' of scalar (discussed in section 3) and vector 
fields on a brane in flat higher dimensional space. 
In these cases, the phenomenological subtleties
which were present for gravitons are gone. As a result, 
one is left with theoretically, as well as 
phenomenologically consistent models of localization for 
spin-0 and spin-1 fields.

\vspace{0.2cm}
{\bf Acknowledgments}
\vspace{0.1cm} \\

The work of G.D. is supported in part by a David and Lucile  
Packard Foundation Fellowship for Science and Engineering. G.G. is
supported by  NSF grant PHY-94-23002. 
M.P. is supported in part by NSF grant PHY-9722083.


\begin{thebibliography}{99}

\bibitem{ADD} N. Arkani-Hamed, S. Dimopoulos, G. Dvali,
Phys. Lett. {\bf B429}, 263 (1998); 
Phys. Rev. {\bf D59}, 0860 (1999); I. Antoniadis,
N. Arkani-Hamed, S. Dimopoulos, G. Dvali, Phys. Lett. 
{\bf B436}, 257 (1998). 

\bibitem{RandallSundrum} L. Randall, R.  Sundrum,
Phys. Rev. Lett. {\bf 83}, 3370  (1999)~; 
Phys. Rev. Lett. {\bf83}, 4690 (1999)~.

\bibitem{DGP1}  G. Dvali, G. Gabadadze, M. Porrati, [hep-th/0002190]~.

\bibitem{Witten} E. Witten, [hep-ph/0002297]~.

\bibitem{DvaliShifman}  
G. Dvali, M. Shifman, Nucl. Phys. {\bf B504}, 127  (1997). 


\bibitem{Dvali} G. Dvali, [hep-th/0004057]~.

\bibitem{GRS1} R. Gregory, V.A.  Rubakov, S.M. Sibiryakov,
[hep-th/0002072]~.

\bibitem{Csaki1} C. Cs\'aki, J. Erlich, T.J. Hollowood, 
[hep-th/0002161]~.





\bibitem{DGP2} G. Dvali, G. Gabadadze, M. Porrati, [hep-th/0003054]~.

\bibitem{Kogan} I. Kogan, G. G. Ross, [hep-th/0003074]~. 

\bibitem{Kang} G. Kang, Y.S. Myung, [hep-th/0003162]~. 

\bibitem{Riccardo} L. Pilo, R. Rattazzi, A. Zaffaroni, [hep-th/0004028]~. 

\bibitem{Csaki2} C. Cs\'aki, J. Erlich, T.J. Hollowood, [hep-th/0003020]~.

\bibitem{GRS2}  R. Gregory, V.A.  Rubakov, S.M. Sibiryakov,
[hep-th/0003045]~.

\bibitem{Capper}
D.M. Capper, Nuovo Cim. {\bf A25} 29 (1975)~. 

\bibitem{Adler} 
S. L. Adler, Phys. Rev. Lett. {\bf 44} 1567 (1980)~; 
Phys. Lett. {\bf B95} 241 (1980)~; 
Rev. Mod. Phys. {\bf 54} 729 (1982); Erratum-ibid. {\bf 55}
837 (1983)~. 

\bibitem{Zee}
A. Zee, Phys. Rev. Lett. {\bf 48} 295 (1982)~. 
 
\bibitem{Khuri} 
N.N. Khuri, Phys. Rev. Lett. {\bf 49} 513 (1982)~;  
Phys. Rev. {\bf D26} 2664 (1982)~.  

\bibitem{Veltman} H. van Dam, M. Veltman, Nucl. Phys. 
{\bf B22}, 397 (1970)~.

\bibitem{Zakharov} V.I. Zakharov, JETP Lett. {\bf 12}, 312 (1970)~.

\bibitem{GarrigaTanaka}
J. Garriga, T. Tanaka, [hep-th/9911055]~

\bibitem{Giddings} 
S.B. Giddings, E. Katz, L. Randall, [hep-th/0002091]~.


\end{thebibliography}
\end{document}